%% file: ICC_2018_draftv9.tex
\documentclass[final]{IEEEtran}

\usepackage{amsthm,amssymb,graphicx,multirow,amsmath,color,amsfonts}
\usepackage[update,prepend]{epstopdf}
\usepackage[noadjust]{cite}
\usepackage[latin1]{inputenc}
\usepackage{tikz}
\usetikzlibrary{arrows,calc}		
\usepackage{bbm} 
\usepackage{pdfpages}
\usepackage{tabulary}
\usepackage{multirow}
\usepackage{float}
\usepackage{comment}

\include{notation}
\allowdisplaybreaks 
\newcommand{\myeq}[1]{\mathrel{\overset{\makebox[0.07pt]{\mbox{(#1)}}}{=}}}

\include{notationITA}

\graphicspath{{./Fig/}}
\title{Integrated mmWave Access and Backhaul in 5G: Bandwidth Partitioning and Downlink Analysis}
\author{
Chiranjib Saha, Mehrnaz Afshang, and Harpreet S. Dhillon 
\thanks{
The authors are with Wireless@VT, Department of ECE, Virginia Tech, Blacksburg, VA, USA. Email: \{csaha,  mehrnaz,  hdhillon\}@vt.edu. The support of the US National Science Foundation (Grant CNS-1617896) is gratefully acknowledged. 
} }

\let\emptyset\varnothing
\begin{document}

\maketitle
\thispagestyle{empty}
\pagestyle{empty}
\vspace{-1.2cm}
\begin{abstract}
With the increasing network densification, it has become exceedingly difficult to provide traditional fiber backhaul access to each cell site, which is especially true for small cell base stations (SBSs). The increasing maturity of millimeter wave (mmWave) communication has opened up the possibility of providing high-speed wireless backhaul to such cell sites. Since mmWave is also suitable for access links, the third generation partnership project (3GPP) is envisioning an integrated access and backhaul (IAB) architecture for the fifth generation (5G) cellular networks in which the same infrastructure and spectral resources will be used for both access and backhaul. 
 In this paper, we develop an analytical framework for IAB-enabled cellular network using which we provide an accurate characterization of its  downlink rate coverage probability. Using this, we  study the performance of two backhaul bandwidth (BW) partition strategies, (i) {\em equal partition:} when all SBSs obtain equal share of the backhaul BW, and (ii) {\em load-based partition:} when the backhaul BW share of an SBS is proportional to its load. Our analysis shows that depending on the choice of the partition strategy, there exists an optimal split of access and backhaul BW for which the rate coverage is maximized. Further, there exists a critical volume of cell-load (total number of users) beyond which the gains provided by  the IAB-enabled network disappear and its performance converges to that of  the traditional macro-only network with no SBSs.
\end{abstract}
\vspace{-0.9em}
\section{Introduction}
With the exponential rise in data-demand far exceeding the capacity of the traditional macro-only cellular network operating in sub-6 GHz bands, network densification using   mmWave base stations (BSs) is becoming a major driving technology for the 5G wireless evolution. While heterogeneous cellular networks (HetNets) with low power SBSs overlaid with traditional macro BSs  improve the spectral efficiency  of the access link (the link between a user and its serving BS), mmWave communication can further boost the data-rate by offering high bandwidth. That  said,  
the HetNet concept  never really turned into a massive real-time deployment since the existing high-speed optical fiber backhaul network that connects the BSs to the network core is not scalabale to the extent of ultra-densification envisioned for  small cells~\cite{DhillonCaire2015}. However, with recent advancement in mmWave communication with highly directional beamforming, it is possible to   replace the so-called {\em last-mile fibers} for  SBSs by establishing fixed  mmWave backhaul links  between the SBS and the MBS  equipped with fiber backhaul, also known as the anchored BS (ABS), thereby achieving Gigabits per second (Gbps) range data-rate over backhaul links. While mmWave fixed wireless backhaul is targetted to be  a part of the first phase of the commercial roll-out   of 5G~\cite{mmWaveMagazineDahlmamn},  3GPP is exploring a more ambitious solution of IAB  
where the ABSs will use  same spectral  resources and infrastructure of mmWave transmission to serve  cellular users in access as well as the SBSs in backhaul~\cite{accessbackhaul3gpp}.
 
Over recent years, stochastic geometry has emerged as a powerful tool for the analysis of cellular networks by modeling the BS and user locations as independent Poisson point processes (PPPs) over an infinite plane. This PPP-based model, initially developed for sub-6 GHz networks~\cite{dhillon2012modeling} has been further tailored for mmWave systems~\cite{MillimeterWaveBai2015,mmWaveHetNetTurgut,AndrewsMMWave}. All these works mostly focus on modeling the access links by including particulars of  channel statistics (e.g. blocking) and transmission techniques (e.g. beamforming) with the assumption of  unconstrained backhaul of mmWave BSs. 
However,  
 there exists no significant prior art on  the analytical framework for IAB-enabled mmWave HetNet except an extension of the PPP-based model~\cite{SinghKulkarniSelfBackhaul}, where the authors model wired and wirelessly backhauled BSs as two independent PPPs. 
 Due to the PPP assumption, this model is unable to capture some  real world morphologies, such as the formation of user {\em hotspots} (or user clusters), which is  one of the key motivations for the deployment of more short-range mmWave SBSs at the locations of these user clusters~\cite{Saha_J1}. Not surprisingly, such configurations are at the heart of the 3GPP simulation models~\cite{saha20173gpp}. 
 To address this shortcoming of the analytical models,  
in this paper, we propose the {\em first 3GPP-inspired} {\em stochastic geometry-based} finite network model for the performance analysis of HetNets with IAB.  The key contributions are summarized next. 

{\em Contributions.} We develop a realistic analytical framework to study the performance of IAB-enabled mmWave HetNets. 
Similar to the models used in 3GPP-compliant simulators~\cite{accessbackhaul3gpp}, the users are assumed to be non-uniformly distributed over the macrocell forming hotspots and the SBSs are located at the geographical centers of the user hotspots. Using this model, we evaluate the downlink rate coverage probability for two backhaul  BW partition strategies, namely, (i) {\em equal partition} where each SBS gets equal share of  BW irrespective of its load\footnote{Throughout the paper, BS-{\em load} refers to the number of users connected to that BS.} and (ii) {\em load-based partition}, where the backhaul BW received by an SBS is proportional to its load.
Three key take-aways of our analysis are: (i) the load-based partition always provides higher rate coverage than the equal partition, (ii) depending on the backhaul BW partition strategy, there exists an optimal access-backhaul BW split for which the rate coverage probability is maximized, and (iii) for  given infrastructure and spectral resources, the IAB-enabled network  outperforms the macro-only network with no SBSs up to a critical volume of total cell-load, beyond which the  performance gains disappear and its performance  converges to that of the macro-only network.
\vspace{-.5em}
\section{System Model}
 \subsection{mmWave Cellular System Model}
\begin{figure}
          \centering
              \includegraphics[width=.87\linewidth]{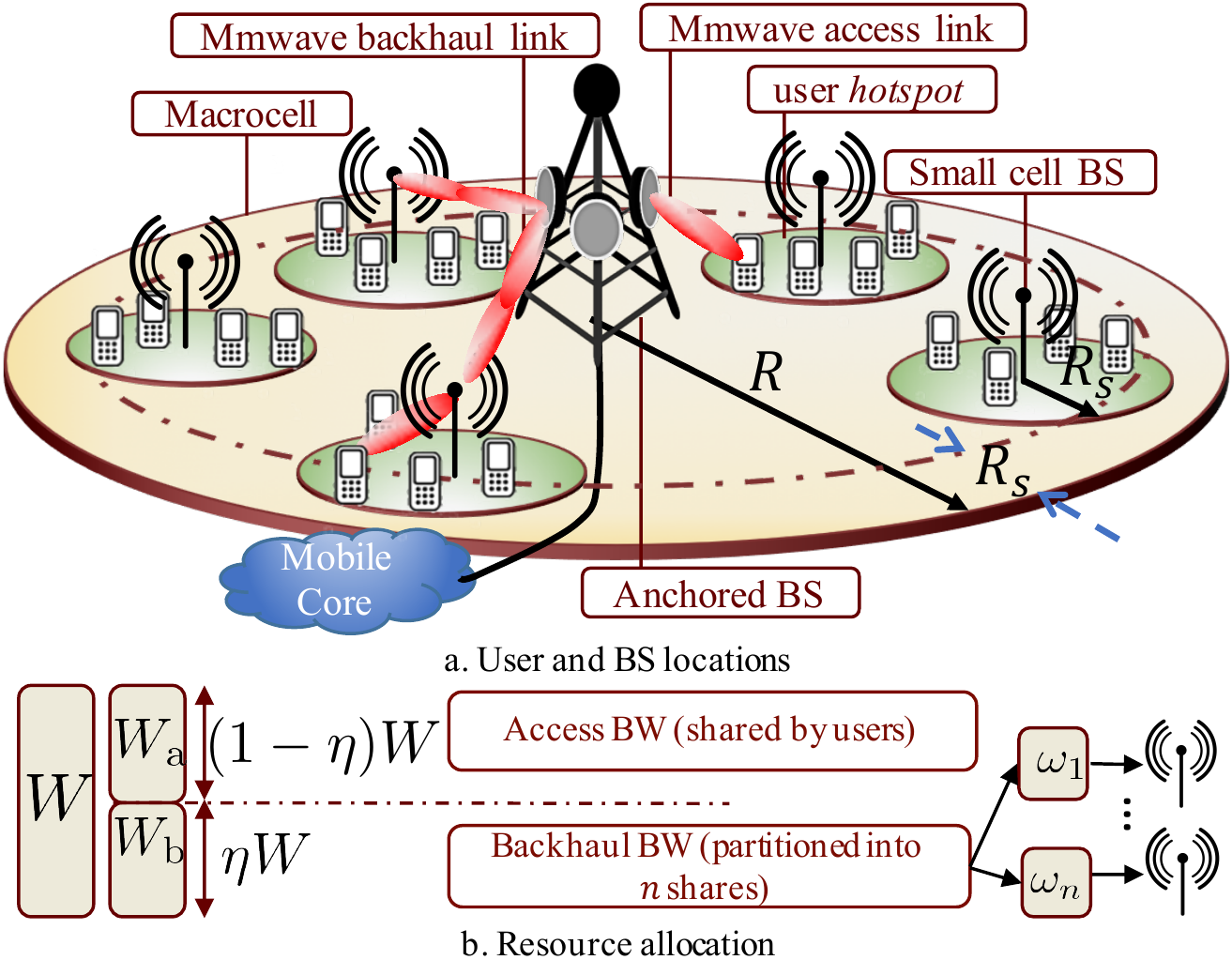}
             \caption{Illustration of system model.}\label{fig::system::model}
          \end{figure}   
\subsubsection{BS and user locations} Inspired by the spatial configurations used in 3GPP simulators~\cite{accessbackhaul3gpp,saha20173gpp} for an outdoor deployment scenario of a two-tier HetNet, we assume  that $n$ SBSs are deployed inside a  circular macrocell of radius $R$ (denoted by $b({\bf 0},R)$) with the macro BS at its center. This BS is called  an ABS since it is connected to the core network with high speed optical fiber. Note that, in contrast to  the  infinite network models (e.g. the PPP-based networks defined over $\R^2$) which are suitable for interference-dominated networks (such as conventional cellular networks in sub-6 GHz), finite network model is a sufficient assumption for  mmWave  networks which are predominantly noise-limited~\cite{Kulkarni_backhaul_asilomar}. 
 The users are assumed to form spatial clusters or hotspots. We model a user hotspot at ${\bf x}$ as  $b({\bf x},R_{\rm s})$, i.e., a circle of radius $R_{\rm s}$ centered at ${\bf x}$. We assume that the macrocell contains $n$ user hotspots,  located at $\{{\bf x}_i\equiv (x_i,\varphi_i), i=1,\dots,n\}$, which are distributed uniformly at random in $b({\bf 0},R-R_{\rm s})$.\footnote{For notational simplicity, we use $x\equiv\|{\bf x}\|,\ \forall\ {\bf x}\in\nbbR^2$.} Thus, $\{{\bf x}_i\}$ is a sequence of independently and identically distributed (i.i.d.) random vectors with the joint distribution of $(x_i,\varphi_i)$ being:
\begin{align}\label{eq::sbs::distribution}
f_{\bf X}({\bf x}_i)=\begin{cases}\frac{x_i}{\pi(R-R_{\rm s})^2}, &\text{when }0<x_i\leq R- R_{\rm s}, \\&0<\varphi_i\leq 2\pi,\\
0, &\text{otherwise.}
\end{cases}
\end{align}The marginal probability density function (PDF) of $x_i$ is obtained as: $f_{X}(x_i)=2x_i/(R-R_{\rm s})^2$ for $0<x_i\leq R- R_{\rm s}$ and $\varphi_{i}$ is a uniform random variable in $(0,2\pi]$.
Note that, this construction  ensures that all hotspots lie entirely  inside the macrocell, i.e., $b({\bf x}_i,R_{\rm s})\cap b({\bf 0},R)^c = \emptyset,\ \forall\ i$. 
For the sake of simplicity of analysis,   we  assume that each hotspot contains a fixed number of users which is equal to $\bar{m}$. 
These $\bar{m}$ users are assumed to be located {\em uniformly at random} in each hotspot. Thus, the location of a user belonging to the hotspot at ${\bf x}_i$ is denoted by ${\bf x}_i+{\bf u}$, where ${\bf u} \equiv (u,\xi)$ is a random vector in $\R^2$ with PDF: 
\begin{align}\label{eq::users::distribution}
f_{\bf U}({\bf u})=\begin{cases}\frac{u}{\pi R_{\rm s}^2}, &\text{when }0< {u}\leq R_{\rm s}, 0<\xi\leq 2\pi\\
0, &\text{otherwise.}
\end{cases}
\end{align} The marginal PDF of $u$ is: $f_{U}(u)=2u/R_{\rm s}^2$ for $0<u\leq R_{\rm s}$ and $\xi$ is a uniform random variable in $(0,2\pi]$.    We assume that the SBSs are deployed at the center of user hotspots, i.e., at $\{{\bf x}_i\}$. The ABS provides wireless backhaul to these SBSs over  mmWave links. For analysis,  we first choose a user uniformly at random along with its hotspot center. We call this hotspot and user (located at ${\bf x}$ and ${\bf x}+{\bf u}$, where, ${\bf x}={\bf x}_n$, without loss of generality)  the {\em representative hotspot} and the  \textit{typical user}, respectively.  See Fig.~\ref{fig::system::model} for an illustration. 
\subsubsection{Propagation assumptions}
 All backhaul and access transmissions are assumed to be performed in mmWave spectrum. We assume that a BS at ${\bf y}\in\nbbR^2$ transmits at  a constant power spectral density (PSD) $P/W$ over a system BW $W$, where  $P = P_{\rm m}$ and $P_{\rm s}$ are the transmission powers of the ABS and SBS,  respectively.   The received power at ${\bf z}$ is given by $P \psi h L({\bf z},{\bf y})^{-1}$, where $\psi$ is the combined antenna gain of the transmitter and receiver and $L ({\rm dB})= \beta (\mathrm{dB})+10\alpha\log_{10}\|{\bf z}-{\bf y}\|$ is the associated path loss.  We assume that all links undergo i.i.d. Nakagami-$m$ fading. Thus, $h\sim{\tt Gamma}(m,m^{-1})$.

\subsubsection{Blockage model}
%
Since mmWave signals are sensitive to physical blockages such as buildings, trees and even human bodies,   the LOS and NLOS path-loss characteristics have to be explicitly considered into analysis.   
On similar lines of \cite{Bai_mmWave}, we assume exponential blocking model.  Each mmWave link of distance $r$ between the transmitter (ABS/SBS) and receiver (SBS/user) is LOS or NLOS according to an independent Bernoulli random variable with LOS probability $p(r) =\exp(-r/\mu)$, where $\mu$ is the LOS range constant depending on the geometry and density of blockages. 

We assume that all BSs are equipped with steerable directional antennas and the user equipments have omni-directional antenna. Let $G$ is the directivity gain of the transmitting and receiving antennas of the BSs (ABS and SBS). Assuming perfect beam alignment, the effective gains on  backhaul and  access links are  $G^2$ and $G$, respectively. We assume that the system is noise-limited, i.e., at any receiver, the  interference is negligible compared to the thermal noise  with PSD ${\tt N}_0$.   Noticing that, the signal-to-noise-ratio ($\snr$) on a link with BW ${W}'$  is proportional to $\frac{\frac{P}{W}{W}'}{{\tt N}_0{W}'} = P/{\tt N}_0W$, the $\snr$-s of a backhaul link   from ABS to SBS at ${\bf x}$,  access links from SBS at ${\bf x}$ to user at ${\bf x}+{\bf u}$ and ABS to user at ${\bf x}+{\bf u}$  are respectively expressed as:
\begin{subequations}
\begin{alignat}{3}
&\snr_{\rm b}({\bf x}) = \frac{P_{\rm m} G^2 h_{\rm b}L({\bf 0},{\bf x})^{-1}}{{\tt N}_0 W},\\ 
&\snr_{\rm a}^{\rm SBS}({\bf x}+{\bf u}) =  \frac{P_{\rm s}G h_{\rm s}L({\bf x},{\bf x}+{\bf u})^{-1}}{{\tt N}_0W},\\
&\snr_{\rm a}^{\rm ABS}({\bf x}+{\bf u}) =  \frac{P_{\rm m}
G h_{\rm m}L({\bf 0},{\bf x}+{\bf u})^{-1}}{{\tt N}_0W},
\end{alignat}
\end{subequations}
where $\{h_{\rm b}, h_{\rm s}, h_{\rm m}\}$ are the corresponding small-scale fading gains.
\subsubsection{User association}
We assume that the SBSs operate in closed-access, i.e., users in hotspot can only connect to the SBS at hotspot center, or the ABS.  Given the complexity of user association in mmWave using beam sweeping techniques, we assume a much simpler way of the user association which is performed by signaling in sub-6 GHz which is analogous to the current LTE standard~\cite{HeathAlkhateeb2017BeamAssociation}. In particular, the BSs broadcast paging signal using omnidirectional antennas in sub-6 GHz and the user associates to the candidate serving BS based on the maximum received power over the paging signals.  
 Since the broadcast signaling is in sub-6 GHz, we   assume  same power-law pathloss function for both LOS and NLOS components  with path-loss exponent $\alpha$ due to rich scattering environment. 
 We define the association event for the typical user  ${\cal E}$ as:
 \begin{align}
 {\cal E} = \begin{cases}
 1 &\text{ if } P_{\rm s}\|{\bf u}\|^{-\alpha} >P_{\rm m}\|{\bf x}+{\bf u}\|^{-\alpha}, \\
  0, &\text{ otherwise,}
 \end{cases}
 \end{align}
 where $\{0,1\}$ denote association to ABS and SBS,  respectively. The typical user at ${\bf x}+{\bf u}$ is  {\em under coverage} in the downlink if either of the following two events occurs:
 \begin{align}
  & {\cal E}  = 1 \text{ and }  \snr_{\rm b}({\bf x})>\theta_1, \snr_{\rm a}^{\rm SBS}({\bf u})>\theta_2, \text{ or,}\notag\\
  &{\cal E}  = 0 \text{ and }\snr_{\rm a}^{\rm ABS}({\bf x}+{\bf u})>\theta_3,\label{eq::coverage::def}
 \end{align}
 where $\{\theta_1, \theta_2, \theta_3\}$ are  the coverage thresholds for successful demodulation and decoding. 
\subsection{Resource allocation} 
The ABS is assumed to be capable of transmitting on both mmWave and sub-6 GHz bands.  The sub-6 GHz band is reserved for control channel and the mmWave band is kept for data-channel. Since mmWave links are unreliable for long distances, a fraction of the macro users  may be offloaded to sub-6 GHz depending on the  link quality from ABS to the user. However, due to space constraints,  we will delegate the discussion on sub-6 GHz offloading to the extended version of this paper.   The total mmWave BW $W$ for downlink transmission is partitioned into two parts, $W_{\rm b}=\eta W$ for backhaul and $W_{\rm a}=(1-\eta)W$ for access, where $\eta\in[0,1)$ determines the access-backhaul split. Each BS is assumed to employ a simple round robin scheduling policy for serving users, under which the total access BW is equally shared among its associated users, referred to  alternatively as {\em load} on that particular BS. On the other hand, we assume that the backhaul BW is shared amongst $n$ SBSs by  either of the following two strategies: 
 1. {\em Equal partition.} This is the  simplest partition strategy  where $W_{\rm b}$ is equally divided into $n$ splits. 
  2. {\em Load-based partition.} In this scheme, the ABS allocates backhaul BW proportional to the load on each small cell.
  If  the SBS at ${\bf x}$ gets backhaul BW  $W_{\rm s}({\bf x})$, then 
\begin{align}\label{eq::bandwidth::partition}
W_{\rm s}({\bf x}) &= \begin{cases} &\frac{W_{\rm b}}{n}, \qquad\qquad\qquad\  \text{for equal partition},\\
&\frac{N^{\rm SBS}_{{\bf x}}}{N^{\rm SBS}_{{\bf x}}+\sum\limits_{i=1}^{n-1}N^{\rm SBS}_{{\bf x}_i}} W_{\rm b}, \text{for load-based partition},
\end{cases}
\end{align}
where $N_{\rm z}^{\rm SBS}$ denotes the load on a SBS at ${\bf z}$. The BW partition is illustrated in Fig.~\ref{fig::system::model}. 
  To compare the performance of these strategies, we first define the network performance metric in the next Section. 
\subsubsection{Downlink data-rate}
The maximum achievable  downlink data-rate, henceforth referred to as simply the {\em data-rate}, on the backhaul link between the ABS and the SBS, the access link between SBS and user, and the access link between ABS and user are  obtained by:
\begin{subequations}
\begin{alignat}{3}
{\cal R}_{\rm b}^{\rm ABS} &=  W_{\rm s}({\bf x})
\log_2(1+\snr_{\rm b}({\bf x})),\label{eq::rate_backhaul}\\
{\cal R}_{\rm a}^{\rm SBS} &= \min\bigg(\frac{W_{\rm a}}{N_{\bf x}^{\rm SBS}}\log_2(1+\snr_{\rm a}^{\rm SBS}({\bf u})), \notag\\
&\frac{ W_{\rm s}({\bf x})}{N_{\bf x}^{\rm SBS}}\log_2(1+\snr_{\rm b}({\bf x}))\bigg),\label{eq::rate_sbs_access} \\
{\cal R}_{\rm a}^{\rm ABS} &= \frac{W_{\rm a} }{N_{\bf x}^{\rm ABS}+\sum\limits_{i=1}^{n-1}N_{{\bf x}_i}^{\rm ABS}}\log_2(1+\snr_{\rm a}^{\rm ABS}({\bf x}+{\bf u})),
\label{eq::rate_abs_access}
\end{alignat}
\end{subequations}
where ${W_{\rm s}}({\bf x})$ is defined  according to backhaul BW  partition strategies in \eqref{eq::bandwidth::partition} and  $N_{\bf z}^{\rm ABS}$ denotes the load on the ABS due to the macro users of the hotspot at ${\bf z}$. In \eqref{eq::rate_sbs_access}, the first  term inside the $\min$-operation is the data-rate achieved under no backhaul constraint  when the access BW $W_{\rm a}$ is equally partitioned within $N_{\bf x}^{\rm SBS}$ users. However, due to finite backhaul,  ${\cal R}_{\rm a}^{\rm SBS}$ is limited by the second term.
\section{Rate Coverage Probability Analysis}
In this Section, we derive the expression of  rate coverage probability  conditioned on ${\bf x}$ and ${\bf u}$ and later decondition over them. This deconditioning averages out all the spatial randomness of the user and hotspot locations in the given network configuration.
\subsection{Association Probability}
The following Lemma characterizes the association probability to the SBS. 
\begin{lemma}\label{lemm::association::sbs}
 Conditioned on the fact that the user belongs to the  hotspot  at ${\bf x}$, the association probability to SBS is given by: ${\cal A}_{\rm s}({\bf x}) ={\cal A}_{\rm s}(x)= $
\begin{align}\label{eq::association::sbs}
\int_0^{2\pi}\frac{\bigg(\min(R_{\rm s},x\frac{k_p\sqrt{(1-k_p^2.\sin^2\xi)+k_p\cos\xi}}{1-k_p^2})\bigg)^2}{R_{\rm s}^2}{\rm d}\xi,
\end{align}
where $k_p = (P_{\rm s}/P_{\rm m})^{1/\alpha}$,
and the association probability to the ABS is given by ${\cal A}_{\rm m}({ x})= 1-{\cal A}_{\rm s}({x})$.
\end{lemma}
\begin{IEEEproof}
Conditioned on the location of the hotspot center at ${\bf x}$,  ${\cal A}_{\rm s}({\bf x}) =$
\begin{align*}
&\nbbP({\cal E} = 1|{\bf x})=\nbbE[{\bf 1}(P_{\rm m}\|{\bf x}+{\bf u}\|^{-\alpha}<P_{\rm s}\|{\bf u}\|^{-\alpha})|{\bf x}]\\
&=\nbbP(P_{\rm m}(x^2+u^2+2xu\cos\xi)^{-\alpha/2}<P_{\rm s}u^{-\alpha})|x)\\&=
\nbbP(u^2(1-k_p^2)-2x\cos\xi k_p^2u-k_p^2x^2<0|x)\\&\myeq{a}\nbbP\bigg(u\in\big(0, \scalebox{0.9}{$\frac{xk_p\sqrt{(1-k_p^2\sin^2\xi)+k_p\cos\xi}}{1-k_p^2}$}\big), \xi\in(0, 2\pi]\big|x\bigg)\\&
\myeq{b}\nbbE\bigg[\frac{\min\bigg(R_{\rm s},\frac{xk_p\sqrt{1-k_p^2\sin^2\xi+k_p\cos\xi}}{1-k_p^2}\bigg)}{R_{\rm s}^2}\big| x\bigg], 
\end{align*}
where $\xi = \arg({\bf u}-{\bf x})$ and is uniformly distributed in $(0,2\pi]$. Here, (a) follows from solving the quadratic inequality inside the indicator function,  (b) follows from deconditioning over ${u}$.
The final form is obtained by deconditioning over $\xi$. 
%
\end{IEEEproof}
We now evaluate the coverage probability of a typical user which is the probability of the occurrence of the events defined in \eqref{eq::coverage::def}.  
\begin{theorem}[Coverage probability]\label{thm::coverage::probability}The coverage probability
 is given by:
 \begin{equation}\label{eq::coverage::probability}
 \pc =\int\limits_{0}^{R-R_{\rm s}}\big(\pc_{\rm s}(\theta_1,\theta_2|x)+\pc_{\rm m}(\theta_3|x)\big)f_X(x){\rm d}x,
 \end{equation}
 where $\pc_{\rm s} (\theta_1,\theta_2|x) =$
 \begin{multline*}
  \int\limits_{0}^{2\pi}\int\limits_{0}^{u_{\max}(x,\xi)} 
\bigg(p({x}) 
F_h\bigg(\frac{ x^{\alpha_L}\beta{\tt N}_0W\theta_1}{P_{\rm m}G^2 },m_L\bigg) +(1-p({x})) \\F_h\bigg(\frac{ x^{\alpha_{NL}}\beta{\tt N}_0W\theta_1}{P_{\rm m}G^2 },m_{NL}\bigg)\bigg)
\bigg(p({u})
F_h\bigg(\frac{ u^{\alpha_L}\beta{\tt N}_0W\theta_2}{P_{\rm s}G },m_{L}\bigg)\\ +(1-p({u}))F_h\bigg(\frac{ u^{\alpha_{NL}}\beta{\tt N}_0W\theta_2}{P_{\rm s} G},m_{NL}\bigg)\frac{f_{ U}(u)}{2\pi}{\rm d}{u}\:{\rm d}{\xi}\bigg),
\end{multline*}
where 
  $u_{\max}(x,\xi) =  \min\bigg(R_{\rm s}, x k_p\frac{\sqrt{(1-k_p^2\sin^2\xi)}+k_p\cos\xi)}{1-k_p^2}\bigg)$ and $F_{h}(\cdot)$ is the {complementary cumulative distribution function (CCDF) of Gamma distribution}, 
    and $\pc_{\rm m} (\theta_3|x)=$
\begin{multline*}
 \int\limits_{0}^{2\pi}\int\limits_{u_{\max}(x,\xi)}^{R_{\rm s}} \scalebox{.95}{$\bigg( p(\kappa(x,u,\xi)) 
F_h\bigg(\frac{{\kappa(x,u,\xi)}^{\alpha_L}\beta{\tt N}_0W\theta_3}{P_{\rm m}G },m_L\bigg)$}\\ +
(1-p(\kappa(x,u,\xi))) F_h\scalebox{0.95}{$\bigg(\frac{ {\kappa(x,u,\xi)}^{\alpha_{NL}}\beta{\tt N}_0W\theta_3}{P_{\rm m}G },m_{NL}\bigg)\bigg)\frac{f_{ U}(u){\rm d}u\:{\rm d}\xi}{2\pi}$},
\end{multline*}
\end{theorem}
where $\kappa(x,u,\xi)=({x^2+u^2+2 x u \cos\xi})^{1/2}$.
\begin{IEEEproof}
%
See Appendix~\ref{app::coverage::probability}.
\end{IEEEproof}
As expected, coverage probability  is the summation of two terms, each corresponding to the probability of occurrences of the two mutually exclusive events appearing in \eqref{eq::coverage::def}. 
\subsection{Load distributions}\label{subsec::load::dist}
While the load distributions for the PPP-based models are well-understood~\cite{KulkarniGhoshAndrews2016}, they are not directly applicable to the 3GPP-inspired finite model used in this paper. Consequently, in this Section, we provide a novel approach to characterize the ABS and SBS load for this model. 
  As we saw in \eqref{eq::rate_abs_access}, the load on the ABS has  two components, one is due to the contribution of the number of users of the {representative} hotspot connecting to the 
ABS (denoted by $N_{{\bf x}}^{\rm ABS}$)  and another is due to the macro users of other clusters, which we lump into a single random variable, $N_{\rm o}^{\rm ABS} = \sum_{i=1}^{n-1}N_{{\bf x}_i}^{\rm ABS}$. On the other hand, $N_{\bf x}^{\rm SBS}$ and $N_{\rm o}^{\rm SBS}=\sum_{i=1}^{n-1}N_{{\bf x}_i}^{\rm SBS}$ denote the load on the SBS at ${\bf x}$ and sum load of all SBSs except the one at ${\bf x}$. 
 First, we obtain the probability mass functions (PMFs) of $N_{{\bf x}}^{\rm ABS}$ and $N_{{\bf x}}^{\rm SBS}$  in the following Lemma. 
\begin{lemma}\label{lemm::load::characterization::abs}
Given the fact that the typical user belongs to a hotspot at ${\bf x}$, load on the ABS  due to the macro users in the hotspot at ${\bf x}$ ($N^{\rm ABS}_{{\bf x}}$) and load on the SBS at $\bf x$  ($N^{\rm SBS}_{\bf x}$) are distributed as follows:
\begin{align}
&\nbbP(N^{\rm ABS}_{{\bf x}}=k|{\bf x})= {\bar{m}-1\choose k-1}{\cal A}_{\rm m}(x)^{k-1}{\cal A}_{\rm s}(x)^{\bar{m}-k}\label{eq::load::abs::load_x::fixedN},\\
&\nbbP(N^{\rm SBS}_{{\bf x}}=k|{\bf x})= {\bar{m}-1\choose k-1}{\cal A}_{\rm s}(x)^{n-1}{\cal A}_{\rm m}(x)^{\bar{m}-k}.\label{eq::load::sbs::load_x::fixedN}
\end{align}
\end{lemma}
\begin{IEEEproof}
Conditioned on the fact that the typical user belonging to the hotspot at $\bf x$ connects to the ABS, $N_{\bf x}^{\rm ABS} =1+ N_{\bf x}^{\rm ABS,o}$, where $N_{\bf x}^{\rm ABS,o} =\nbbE[\sum_{j=1}^{\bar{m}-1} \mathbf{1}({P_{\rm m}\|{\bf x}+{\bf u}_j\|^{-\alpha}>P_{\rm s}\|{\bf u}_j\|^{-\alpha}})|{\bf x}]$, is the load due to  other users among  $\bar{m}-1$ users in the hotspot connecting to the ABS. The conditional moment generating function (MGF) of $ N_{\bf x}^{\rm ABS,o}$ is:  $\nbbE[e^{sN_{\bf x}^{\rm ABS,o}}|{\bf x}] =$
\begin{align*} 
&\nbbE\bigg[\prod\limits_{j=1}^{\bar{m}-1}e^{s\mathbf{1}({P_{\rm m}\|{\bf x}+{\bf u}_j\|^{-\alpha}>P_{\rm s}\|{\bf u}_j\|^{-\alpha}})}|{\bf x}\bigg] 
\\&\myeq{a}\prod\limits_{j=1}^{\bar{m}-1}\nbbE[e^{s\mathbf{1}({P_{\rm m}\|{\bf x}+{\bf u}_j\|^{-\alpha}>P_{\rm s}\|{\bf u}_j\|^{-\alpha}})}|{\bf x}]\\&= \prod\limits_{j=1}^{\bar{m}-1}e^s\nbbP({P_{\rm m}\|{\bf x}+{\bf u}_j\|^{-\alpha}>P_{\rm s}\|{\bf u}_j\|^{-\alpha}}|{\bf x})+\nbbP(P_{\rm s}\|{\bf u}_j^{-\alpha}\|>\\&P_{\rm m}\|{\bf x}+{\bf u}_j\|^{-\alpha}|{\bf x}) = ({\cal A_{\rm m}}(x)e^s + (1-{\cal A_{\rm m}}(x)))^{\bar{m}-1},
\end{align*} 
which is the MGF of a Binomial distribution with $(\bar{m}-1,{\cal A}_{\rm m})$. Here,  (a) follows from the fact that ${\bf u}_j$-s are i.i.d. The PMF of $N_{\bf x}^{\rm SBS}$ can be obtained on similar lines by altering the inequality in the first  step of the above derivation.
\end{IEEEproof}
We now obtain the PMFs of $N_{\rm o}^{\rm ABS}=\sum_{i=1}^{n-1}N_{{\bf x}_i}^{\rm ABS}$ and $N_{\rm o}^{\rm SBS}$  in the following Lemma. {Note that, since ${\bf x}_i$-s are i.i.d., $N_{\rm o}^{\rm ABS}$ and $N_{\rm o}^{\rm SBS}$ are independent of ${\bf x}$.}  In what follows, the exact PMF of $N_{{\rm o}}^{\rm ABS}$ ($N_{{\rm o}}^{\rm SBS}$) is in the form of  $(n-1)$-fold discrete convolution and hence is not computationally efficient beyond very small values of $n$.  We present an alternative easy-to-use expression of this PMF by invoking central limit theorem (CLT). In the numerical Section, we verify that this approximation is {tight} even for moderate   values of $n$.
\begin{lemma}\label{lemm::load::characterization::others}
Given the fact that the typical user belongs to a hotspot at ${\bf x}$, load on the ABS due to all other $n-1$ hotspots   is distributed as:
\begin{align}
&\nbbP(a<N_{\rm o}^{\rm ABS}<b)\to\int\limits_{a}^b \frac{1}{\sigma_{\rm m}\sqrt{2\pi}}e^{-\frac{(t-\upsilon_{\rm m})^2}{\sigma_{\rm m}^2}}{\rm d}t,\ \text{for large $n$,}\notag
\end{align}
and sum of the loads on the other SBSs at  $\{{\bf x}_i, i=1\}$  is distributed as:
\begin{align}
&\nbbP(a<N_{\rm o}^{\rm SBS}<b)\to\int\limits_{a}^b \frac{1}{\sigma_{\rm s}\sqrt{2\pi}}e^{-\frac{(t-\upsilon_{\rm s})^2}{\sigma_{\rm s}^2}}{\rm d}t,\ \text{for large $n$,}\notag
\end{align}
where, 
\begin{equation*}
\upsilon_{\rm m}=(n-1)\bar{m}\nbbE[{\cal A}_{\rm m}(X)], \upsilon_{\rm s}=(n-1)\bar{m}\nbbE[{\cal A}_{\rm s}(X)],
\end{equation*}
and 
\begin{equation*}
\sigma_{\rm m}^2=(n-1)[\bar{m}\nbbE[A_{\rm m}(X)A_{\rm s}(X)]+\bar{m}^2{\rm Var}[{\cal A}_{\rm m}(X)]]=\sigma_{\rm s}^2.
\end{equation*}
Here, $\nbbE[{\cal A}_{\rm m}(X)]=\int_{0}^{R-R_{\rm s}}{\cal A}_{\rm m}(x)f_{X}(x){\rm d}x$, and ${\rm Var}[{\cal A}_{\rm m}(X)] = \int\limits_{0}^{R-R_{\rm s}}\big({\cal A}_{\rm m}(x)\big)^2f_X(x){\rm d}x - (\nbbE[{\cal A}_{\rm m}(X)])^2$.
\end{lemma}
\begin{IEEEproof}
Following the proof of Lemma~\ref{lemm::load::characterization::abs}, conditioned on the location of a hotspot at ${\bf x}_i$, $N_{{\bf x}_i}^{\rm ABS}$ becomes a Binomial random variable with $(\bar{m}, {{\cal A}}_{\rm m}({ x}_i))$. Now, $N_{\rm o}^{\rm ABS } = \sum_{i=1}^{n-1} \nbbE_{{\bf x}_i}[N_{{\bf x}_i}^{\rm ABS }]$, where $\nbbE_{{\bf x}_i}[N_{{\bf x}_i}^{\rm ABS }]$-s are i.i.d. with $\nbbP(\nbbE_{{\bf x}_i}[N_{{\bf x}_i}^{\rm ABS }]=k) = \nbbE_{{ x}_i}[{\bar{m}\choose k}{\cal A}_{\rm m}(x_i)^{k}{\cal A}_{\rm s}(x_i)^{\bar{m}-k}]$.  The exact PMF of $N_{\rm o}^{\rm ABS }$ is obtained by the $(n-1)$-fold discrete convolution of this PMF. We avoid this complexity of the exact analysis by first characterizing the mean and variance of $N_{\rm o}^{\rm ABS}$ as: $\upsilon_{\rm m}= \nbbE[N_{\rm o}^{\rm ABS}] = \sum_{i=1}^{n-1}\nbbE [\nbbE_{{\bf x}_i}[N_{{\bf x}_i}^{\rm ABS }]]=(n-1)\nbbE_{x_i}[\bar{m}{\cal A}_{\rm m}(x_i)]$, and $\sigma_{\rm m}^2 = {\rm Var}[N_{\rm o}^{\rm ABS}]\myeq{a}\sum_{i=1}^{n-1} {\rm Var}[\nbbE_{{\bf x}_i}[N_{{\bf x}_i}^{\rm ABS}]]=\sum_{i=1}^{n-1} \nbbE_{{\bf x}_i}[\nbbE[(N_{{\bf x}_i}^{\rm ABS})^2]]-(\nbbE_{{\bf x}_i}[N_{{\bf x}_i}^{\rm ABS}])^2 =\sum_{i=1}^{n-1} \nbbE_{x_i}[\bar{m}{\cal A}_{\rm m}(x_i){\cal A}_{\rm s}(x_i)+(\bar{m}{\cal A}_{\rm m}(x_i))^2]  -(\bar{m}\nbbE_{x_i}[{\cal A}_{\rm m}(x_i)])^2$, where (a) is due to the fact that $\nbbE_{{\bf x}_i}[N_{{\bf x}_i}^{\rm ABS }]$-s are i.i.d. The final result follows from some algebraic manipulation. Having derived the mean and variance of  $N_{\rm o}^{\rm ABS}$, we invoke CLT to approximate the distribution of $N_{\rm o}^{\rm ABS}$ since it can be represented as a sum of i.i.d. random varables with finite mean and variance. Similar steps can be followed for the distribution of   $N_{\rm o}^{\rm SBS}$. 
\end{IEEEproof}
\subsection{Rate Coverage Probability}
 We first define the downlink rate coverage probability (or simply, rate coverage) as follows. 
\begin{ndef}[Rate coverage probability]\label{def::rate::coverage} The rate coverage probability of a link with BW $\tilde{W}$ is defined as the probability that the maximum achievable data rate (${\cal R}$) exceeds a certain threshold $\rho$, i.e., $\nbbP({\cal R}>\rho)  =$
\begin{align}
 \nbbP\bigg(\tilde{W}\log_{2}(1+\snr)>\rho\bigg) = \nbbP(\snr>2^{{\rho}/{\tilde{W}}}-1).
\end{align}
\end{ndef}
Hence, we see that the rate coverage probability is the coverage probability evaluated at a modified $\snr$-threshold. We now evaluate the rate coverage probability  for different backhaul BW partition strategies in the following Theorem. 
\begin{theorem}\label{thm::rate::cov::equal::partition}The rate coverage probability for a target data-rate $\rho$  is given by:
\begin{equation}
\pr = \pr_{\rm m} + \pr_{\rm s},
\end{equation}
where $\pr_{\rm m}$ ($\pr_{\rm s}$) denotes the ABS-rate coverage (SBS-rate coverage) which is the probability that the typical user is receiving data-rate greater than or equal to $\rho$ and is served by the ABS (SBS). The ABS-rate coverage is given by:
\begin{multline}\label{eq::rate::cov::macro::fixedN}
\pr_{\rm m} = \sum\limits_{k=1}^{\bar{m}}{{\bar{m}-1} \choose k-1}\int\limits_{-\infty}^{\infty}\int\limits_{0}^{R-R_{\rm s}}\pc_{\rm m}\bigg(2^{\frac{\rho(t+k)}{W_{\rm a}}}-1|x\bigg)\times\\{\cal A}_{\rm m}(x)^{k-1}{\cal A}_{\rm s}(x)^{\bar{m}-k} f_X(x){\rm d}x \frac{1}{\sigma_{\rm m}\sqrt{2\pi}}e^{-\frac{(t-\upsilon_{\rm m})^2}{\sigma_{\rm m}^2}}{\rm d}t.
\end{multline}
The SBS-rate coverage depends on the backhaul BW partition strategy. For   equal partition, $\pr_{\rm s} = $
\begin{multline}
\sum\limits_{k = 1}^{\bar{m}}{{\bar{m}-1} \choose k-1}\int\limits_{0}^{R-R_{\rm s}}\pc_{\rm s}\bigg(2^{\frac{\rho n k}{W_{\rm b}}}-1,2^{\frac{\rho k}{W_{\rm a}}}-1\big|x\bigg)\times\\{\cal A}_{\rm s}(x)^{k-1}{\cal A}_{\rm m}(x)^{\bar{m}-k}f_X(x){\rm d}x,\label{eq::rate::cov::sbs::fixedN::eq::partition}
\end{multline}
and for load-based partition, $\pr_{\rm s} =$
\begin{multline}\label{eq::rate::cov::inst-load::sbs::fixedN}
 \sum\limits_{k = 1}^{\bar{m}}{{\bar{m}-1} \choose k-1}\int\limits_{-\infty}^{\infty}\int\limits_{0}^{R-R_{\rm s}}\pc_{\rm s}\bigg(2^{\frac{\rho(k+t)}{W_{\rm b}}}-1,2^{\frac{\rho k}{W_{\rm a}}}-1\big| x\bigg)\\\times{\cal A}_{\rm s}(x)^{k-1}{\cal A}_{\rm m}(x)^{\bar{m}-k}f_{X}(x){\rm d}x \frac{1}{\sigma_{\rm s}\sqrt{2\pi}}e^{-\frac{(t-\upsilon_{\rm s})^2}{\sigma_{\rm s}^2}}{\rm d}t.
\end{multline}
\end{theorem}
\begin{IEEEproof}See Appendix~\ref{app::rate::cov::equal::partition}.
\end{IEEEproof}
\section{Results and Discussion}
\begin{figure}
      \centering
      \begin{minipage}{0.49\linewidth}
          \begin{figure}[H]
              \includegraphics[width=\linewidth]{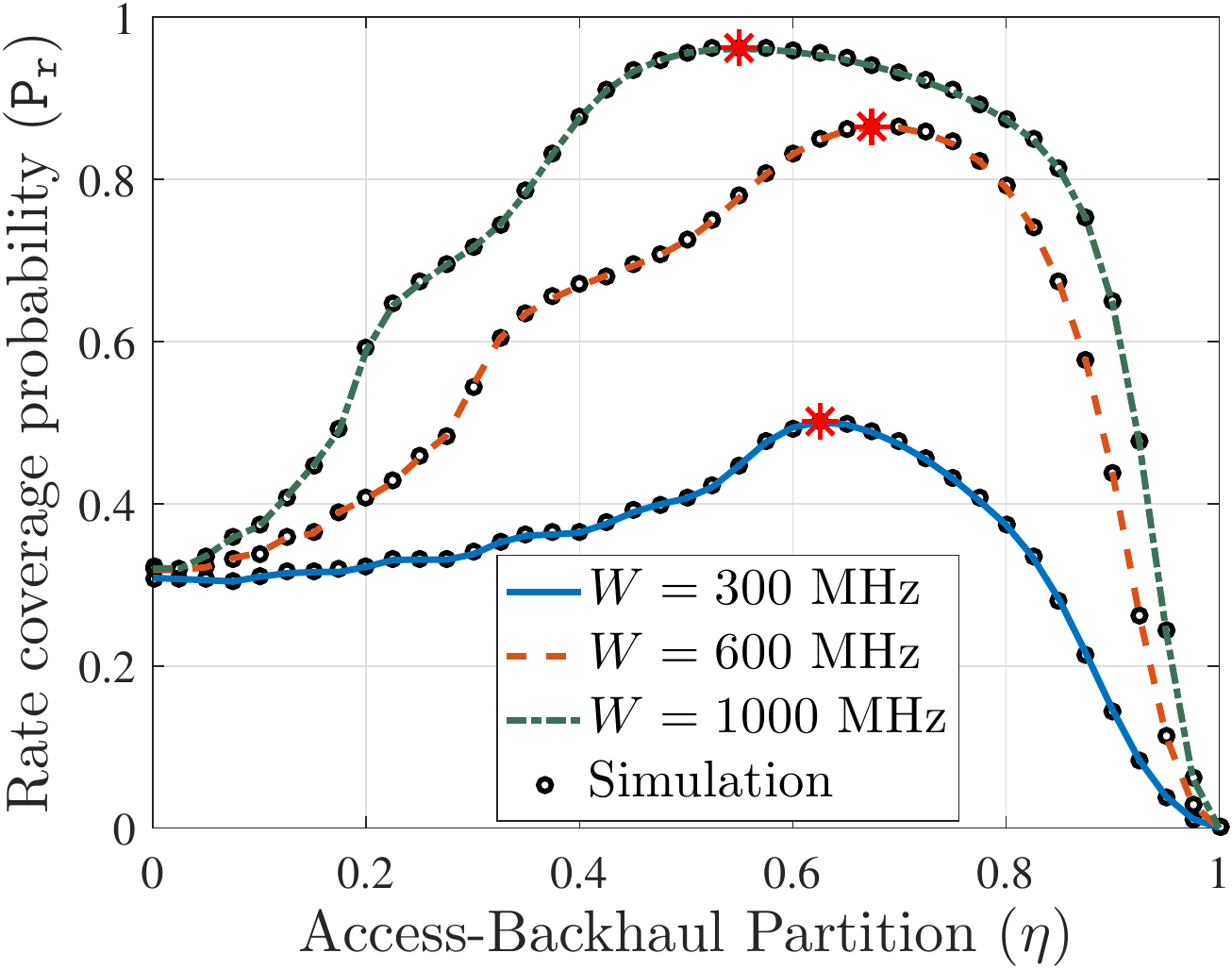}
             \caption{Rate coverage probability for equal-based partition for different bandwidths ($\rho = 50$ Mbps).}\label{fig::comparison::rate::cov::bw::eq}
          \end{figure}
      \end{minipage}
      \begin{minipage}{0.49\linewidth}
          \begin{figure}[H]
              \includegraphics[width=\linewidth]{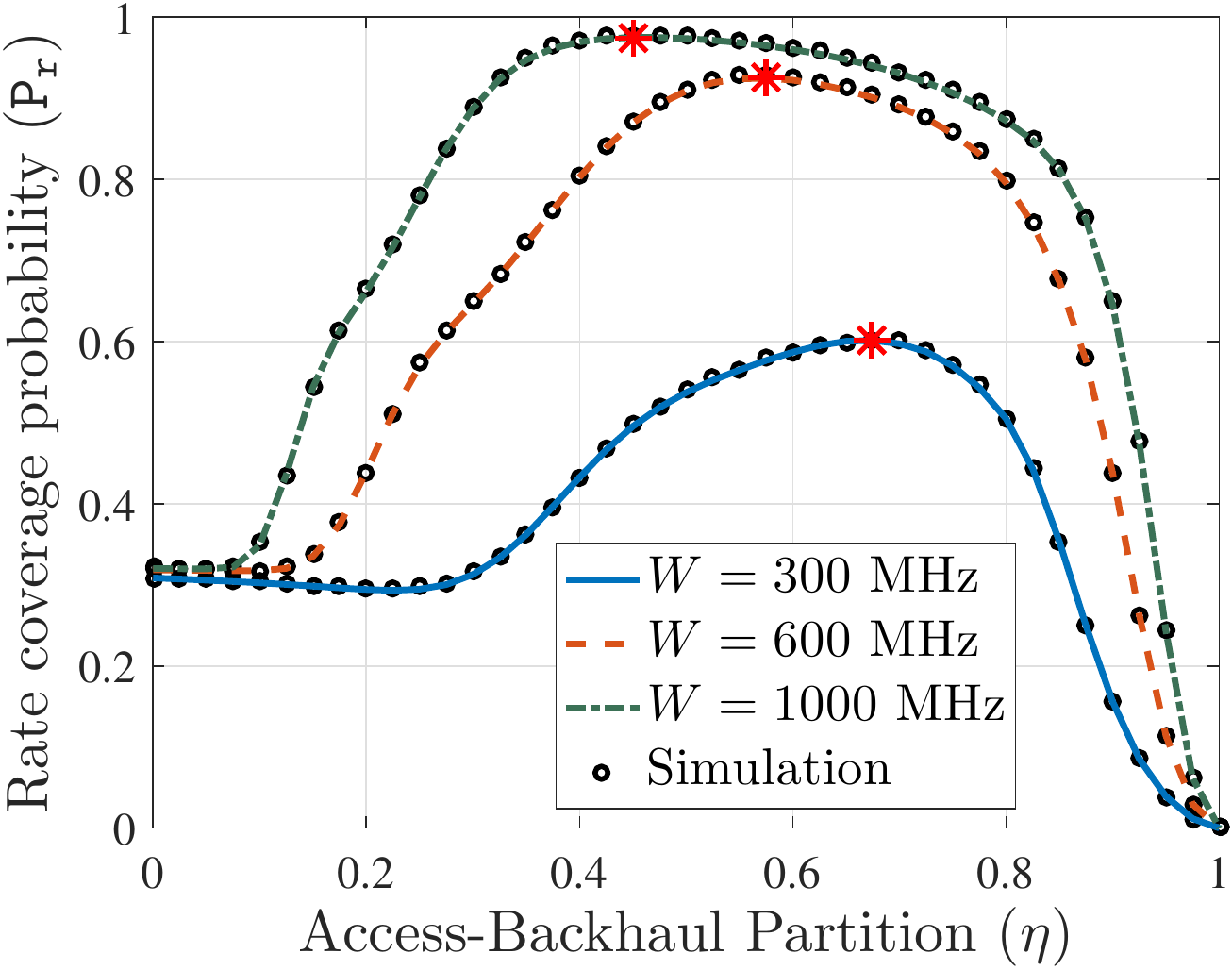}
               \caption{Rate coverage probability for load-based partition for different bandwidths ($\rho = 50$ Mbps).}\label{fig::comparison::rate::cov::bw::load}
          \end{figure}
      \end{minipage}
      \vspace{-1.6em}
            \begin{minipage}{0.49\linewidth}
          \begin{figure}[H]
             \includegraphics[width=\linewidth]{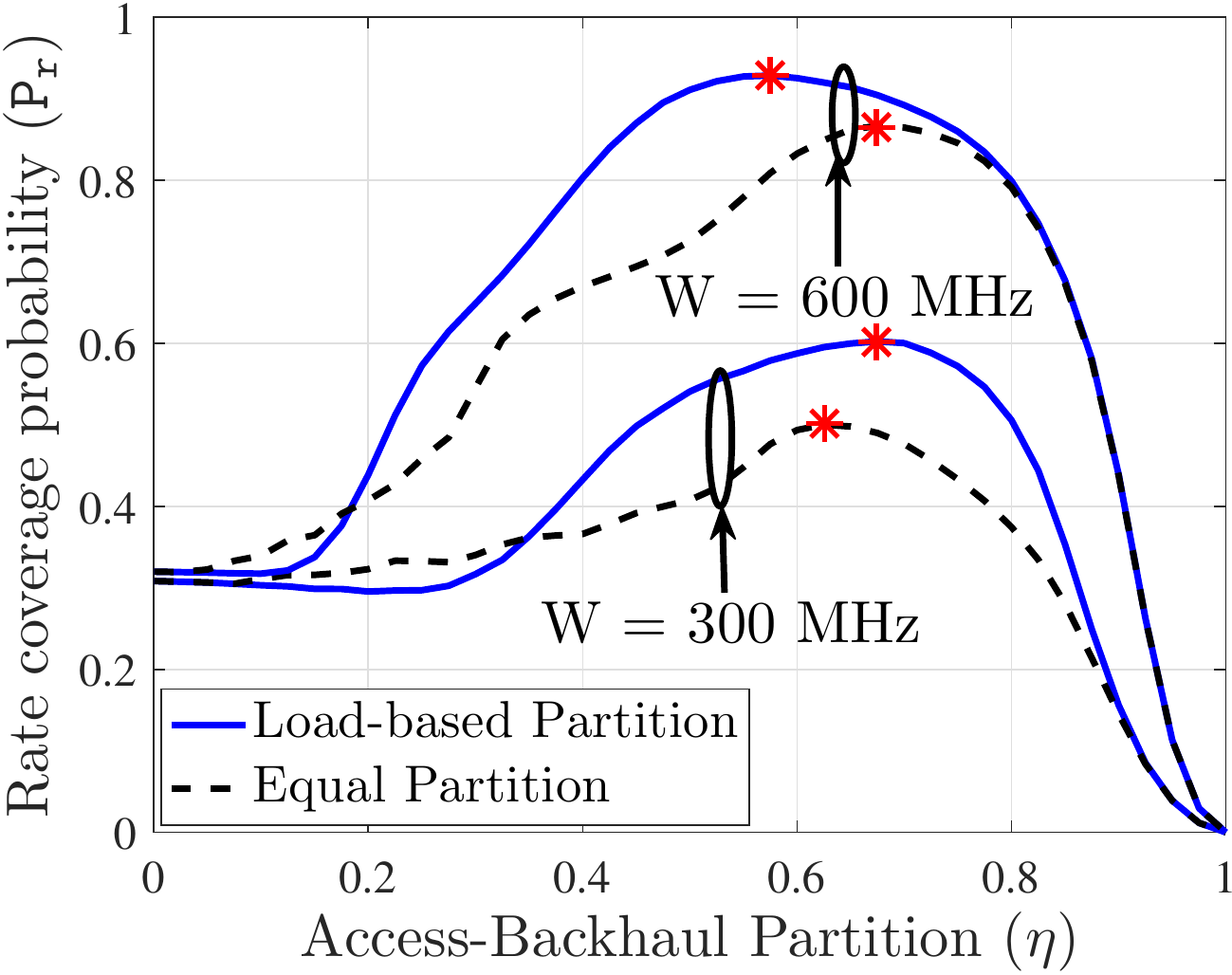}
\caption{Comparision of equal and load-based parition strategies  ($W = 300$ MHz, $\rho = 50$ Mbps).}\label{fig::comparison::rate::cov::bw}

          \end{figure}
      \end{minipage}
      \begin{minipage}{0.49\linewidth}
          \begin{figure}[H]
            \includegraphics[width=\linewidth]{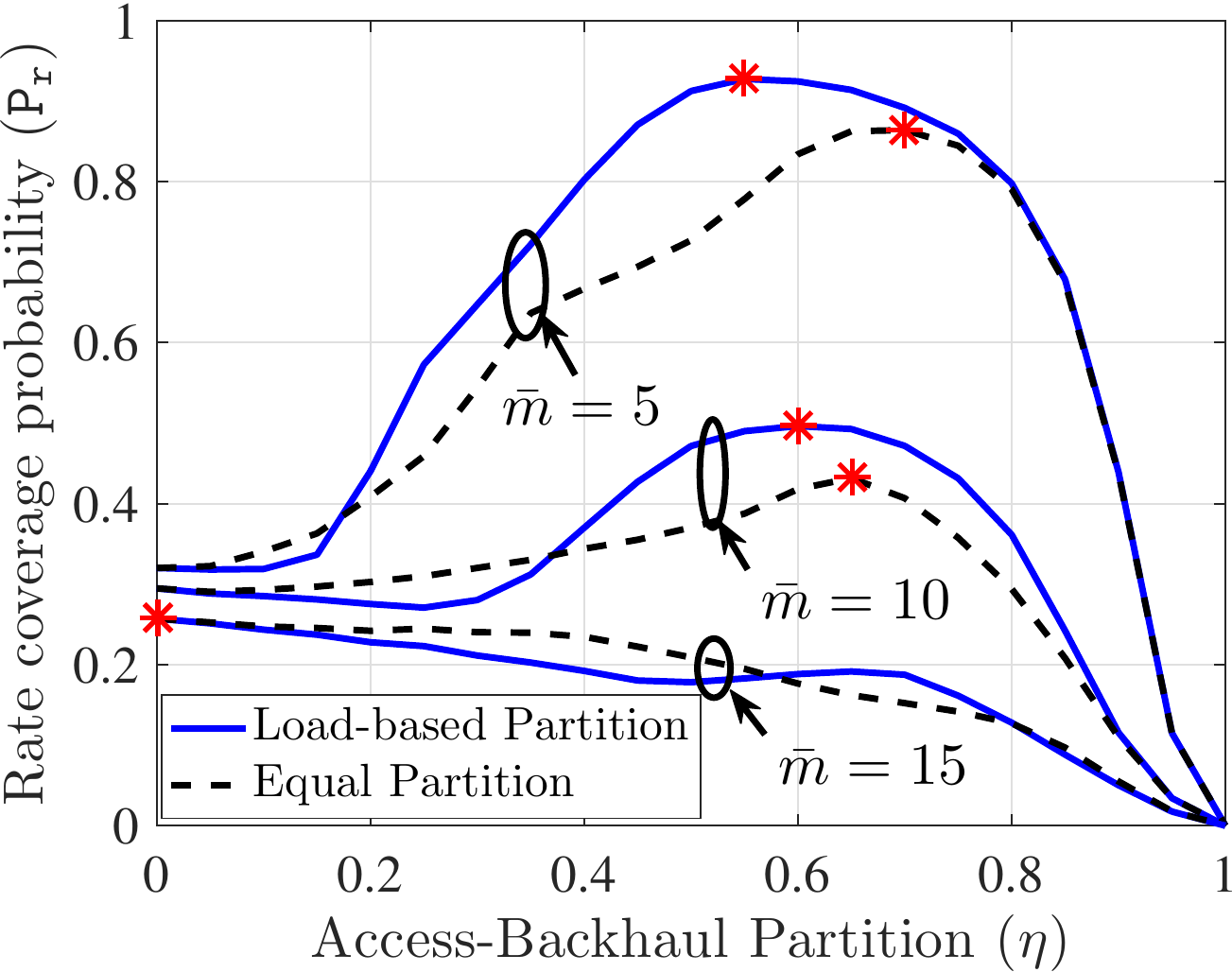}
\caption{Rate coverage for different numbers of users per hotspot ($W = 600$ MHz, $\rho = 50$ Mbps).}\label{fig::comparison::rate::cov::users}
          \end{figure}
      \end{minipage}
  \end{figure}

To validate the results obtained in Theorem~\ref{thm::rate::cov::equal::partition} with simulation, we set the parameters according to Table~\ref{tab::parameters} and plot the values of $\pr$ in Figs~\ref{fig::comparison::rate::cov::bw::eq} and \ref{fig::comparison::rate::cov::bw::load}. 
Recall that one part of ABS and SBS load was approximated using CLT in Lemma~\ref{lemm::load::characterization::others} for efficient computation.  Yet, we obtain a perfect match between simulation and analysis even for $n=10$.  From Fig.~\ref{fig::comparison::rate::cov::bw}, we  observe that, (i) $\pr = 0$ for $\eta=1$ since this corresponds to the extreme when no BW is given to access links, (ii) the rate coverage is maximized  for a particular access-backhaul BW split ($\eta^*=\arg\max_{\{\eta\}}\pr$), and (iii) the maximum rate coverage, $\pr^* = \pr(\eta^*)$ (marked as `*' in Figs~\ref{fig::comparison::rate::cov::bw::eq}-\ref{fig::comparison::rate::cov::users}) in load-based partition is always greater than $\pr^*$ obtained in equal partition. However, note that, load-based partition requires periodic feedback of the load information from the SBSs to the ABS and thus has slightly higher signaling overhead than the equal partition. This opens up an interesting  performance-complexity trade-off for the design of cellular networks with integrated access and backhaul which will be discussed in the extended version of this paper. Next, we plot the effect of increasing system BW on rate coverage. As expected, $\pr$ increases as $W$ increases. However, the increment of $\pr^*$ saturates for very high values of $W$ since high noise power reduces the link spectral efficiency. Another interesting observation is that starting from $\eta = 0$ to $\eta^*$, $\pr$ does not increase monotonically. This is due to the fact that sufficient BW needs to be steered from access to backhaul so that the network with IAB performs better than the macro-only network (corresponding to $\eta = 0$). Finally, we plot the variation of $\pr$ with $\bar{m}$ in Fig.~\ref{fig::comparison::rate::cov::users}. We observe that as $\bar{m}$ increases,   more number of users share the BW and as a result, $\pr$ decreases. However, the optimality of $\pr$ completely disappears for very large value of $\bar{m}$ ($10<\bar{m}<15$ in our case). This implies that for given BW $W$ there exists a critical total cell-load ($n\bar{m}$) beyond which the gain obtained by the IAB architecture  completely disappears.  
\begin{table}
\centering
\caption{{Key system parameters}}
\label{tab::parameters}
\vspace{-1em}
\scalebox{0.8}{
\begin{tabular}{|l|l|l|}
\hline
Notation & Parameter & Value \\ \hline
$P_{\rm m},\ P_{\rm s}$ & mmWave BS transmit powers  & 30, 0 dBm \\ \hline
$\alpha_L, \alpha_{NL}$ & Path-loss exponent &  2.0,  3.3\\ \hline
$\beta$ & Path loss at 1 m & 70 dB \\\hline
$G$ & Main lobe gain & 18 dB \\ \hline
$\mu$ & LOS range constant & 30 m \\ \hline
${\tt N}_0W$ & Noise power & \begin{tabular}[c]{@{}l@{}}-174 dBm/Hz+ $10\log_{10} W$ \\+10 dB {(noise-figure)}\end{tabular} \\ \hline
$m_L,m_{NL}$& Parameter of Nakagami distribution& 2,  3\\\hline
$R$, $R_{\rm s}$& Macrocell and hotspot radius & 40 m, 5m\\
\hline
$n$& Number of  hotspots in macrocell & 10\\\hline
$\rho$& Rate threshold & 50 Mbps\\
\hline
\end{tabular}
}
\end{table}
\vspace{-1em}
\section{Conclusion}
In this paper, we proposed the first 3GPP-inspired  analytical framework for  two-tier mmWave HetNets with IAB and investigated two  backhaul BW partition strategies. Our analysis lead to two important findings:  (i) the existence of the optimal access-backhaul bandwidth partition split, and (ii)  maximum total network load that can be supported using the IAB  architecture. 
This work has numerous extensions such as studying multihop backhauling, impact of beam-alignment errors and cellular  interference on system performance, and more comprehensive performance analysis by considering uplink and downlink transmissions jointly.
\vspace{-1em}
\appendix
\subsection{Proof of Theorem~\ref{thm::coverage::probability}}
\label{app::coverage::probability}
Conditioned on the location of the typical user at ${\bf u}=(u,\xi)$ and its hotspot center at ${\bf x}$, $\pc_{\rm s}(\theta_1,\theta_2|{ x}) =$
\begin{align*}
&  \nbbP(\snr^{\rm SBS}_{\rm a}({ u})>\theta_2, \snr_{\rm b}({ x})>\theta_{1},{\cal E}=1|x)\\
&=\nbbP(\snr^{\rm SBS}_{\rm a}({ u})>\theta_2,{\cal E}=1|x)\nbbP(\snr_{\rm b}({ x})>\theta_{1}|x)\\
&\myeq{a}\nbbE\bigg[\bigg(p({u}) \nbbP\scalebox{0.98}{$\bigg(\frac{P_{\rm s}G\beta^{-1} h_{\rm s(L)} u^{-\alpha_L}}{{\tt N}_0W}>\theta_2\bigg)$} +(1-p({u}))\times\\& \nbbP\scalebox{0.98}{$\bigg(\frac{P_{\rm s}\beta^{-1} Gh_{\rm s(NL)} u^{-\alpha_{NL}}}{{\tt N}_0W}>\theta_2\bigg)$}\bigg){\bf 1}\big(u\in\scalebox{0.86}{$\big(0,{\frac{xk_p\sqrt{1-k_p^2\sin^2\xi+k_p\cos\xi}}{1-k_p^2}}\big)$},\\& \xi\in(0,2\pi]\big)\bigg|x\bigg]
\bigg(p({x}) \nbbP\bigg(\frac{P_{\rm m}\beta^{-1} G^2 h_{\rm b(L)} x^{-\alpha_L}}{{\tt N}_0W}>\theta_1|x\bigg)\\& +(1-p({x}))
\nbbP\bigg(\frac{P_{\rm m}\beta^{-1} G^2h_{\rm b(NL)} x^{-\alpha_{NL}}}{{\tt N}_0W}>\theta_1|x\bigg)\bigg).
\end{align*}
 Here (a) follows from step (a) in the proof of Lemma~\ref{lemm::association::sbs}. The final form is obtained by evaluating the expectation with respect to $u$ and $\xi$. We can similarly obtain $\pc_{\rm m}(\theta_3|{\bf x})=\nbbP(\snr^{\rm ABS}_{\rm a}({ u})>\theta_3, {\cal E}=0|x)$.
\subsection{Proof of Theorem~\ref{thm::rate::cov::equal::partition}}
\label{app::rate::cov::equal::partition}
First we evaluate $\pr_{\rm m}$ as: $\pr_{\rm m} = \nbbP({\cal R}_{\rm a}^{\rm ABS}>\rho)=$
\begin{align*}
&\nbbP\bigg(\frac{W_{\rm a}}{N_{\bf x}^{\rm ABS}+N_{\rm o}^{\rm ABS}}\log_2(1+\snr_{\rm a}^{\rm ABS}({\bf x}+{\bf u}))>\rho\bigg)\\
&=\nbbP\scalebox{0.86}{$\bigg(\snr_{\rm a}^{\rm ABS}({\bf x}+{\bf u})>2^{\frac{\rho(N_{\bf x}^{\rm ABS}+N_{\rm o}^{\rm ABS})}{W_{\rm a}}}-1\bigg)$}\\&= \pc_{\rm m}\scalebox{0.86}{\bigg($2^{\frac{\rho(N_{\bf x}^{\rm ABS}+N_{\rm o}^{\rm ABS})}{W_{\rm a}}}-1\bigg)$},
\end{align*}
where,  the first step follows from \eqref{eq::rate_abs_access}. The final form is obtained by deconditioning with respect to $N_{\bf x}^{\rm ABS}$, $N_{\rm o}^{\rm ABS}$ and $\bf x$.
 On similar lines, we can evaluate $\pr_{\rm s}$ for equal and load-based partitions. The detailed proof is omitted due to space constraints. 
\vspace{-1em}
\bibliographystyle{IEEEtran}
\bibliography{ICC_2018_draftv9.bbl}
\end{document}

%% file: notationITA.tex
\def\nb0{{\mathbf{0}}}
\def\nb1{{\mathbf{1}}}

\def\nbbE{{\mathbb{E}}}

\def\nbbP{{\mathbb{P}}}

\def\nbbR{{\mathbb{R}}}

\newtheorem{lemma}{Lemma}

\newtheorem{ndef}{Definition}

\newtheorem{theorem}{Theorem}

\def\pc{\mathtt{P_c}}
\def\pr{\mathtt{P_r}}

\def\R{\mathbb{R}}

\def\snr{\mathtt{SNR}}

